\documentclass[10pt,prd,aps,amssymb,amsmath,tightenlines]{revtex4}
\usepackage{graphics}
\usepackage[pdftex]{graphicx}

\newcommand{\ket}{\rangle}
\newcommand{\bra}{\langle}

\begin{document}
\preprint{}

\title{Another look at the optimal Bayes cost in the binary decision problem}

\author{Bernhard K. Meister}
\email{b.meister@imperial.ac.uk}
\affiliation{ Department of Physics, Renmin University of China, Beijing, China 100872}

\date{\today }

\begin{abstract}

The problem of quantum state discrimination between two wave functions
of a particle in a square well potential is considered. The optimal
minimum-error probability for the state discrimination is known to be
given by the Helstrom bound. A new strategy is introduced here whereby
the square well is compressed isoenergetically, modifying  the wave-functions.
The new contracted chamber is then probed using the conventional optimal 
strategy, and the error probability is calculated. It is shown that in 
some cases the  Helstrom bound can be violated, i.e. the state discrimination 
can be realized with a smaller error probability.

\end{abstract}
\maketitle


\section{Introduction}
\label{sec:1a}

Experimental design and data analysis are common challenges in science, and particular acute in quantum mechanics, due to recurring questions about foundational issues and restrictions on measurements.
In the literature many different approaches are discussed. For example, there is the information theoretic approach, where one maximizes the mutual information, and the minimax approach, where one minimizes the maximum cost of a set of strategies assuming a perfidious opponent. Both methods are popular, but this paper will instead use the Bayes procedure to minimize the expected cost, since the existence of a {\it prior} associated with the states to be distinguished is assumed.

 Two disparate concepts are combined in the paper: quantum state discrimination and the isoenergetic modification of the  quantum potential.
Quantum state discrimination or more generally called Bayesian hypothesis testing was developed by Helstrom and others \cite{helstrom, holevo, yuen}. The particular problem of quantum state discrimination  between two possible states with given prior and transition probability was also studied in various settings, e.g. for a more recent paper on this topic see Brody {\it et al.} \cite{dbbm96}. It is generally accepted, but will be challenged in the paper, that the optimal Bayes cost in the binary case, is given by the Helstrom bound, which only depends on the prior and the transition probability between the states and can be written in a simple closed form. 
Expansion of a one-dimensional infinite well in a isoenergetic setting, was studied in Bender {\it et al.} \cite{bbm2002}. The authors established that this type of expansion or contraction has various curious effects like shifting the probabilities between different eigenstates. Here we study implications of their results for the distinguishability of states. 

Next a description of the setup and the procedure to be analyzed. With fixed probabilities, given by the prior, one of two quantum states is put into a square well with infinitely high walls.
Two strategies for calculating the Bayes cost are proposed. In the first strategy,  the combination of prior and transition probability between the two quantum states alone is sufficient to calculate the conventional optimal minimum error probability, i.e. the Helstrom bound. This
error probability can be achieved by a well-known measurement strategy \cite{helstrom} not further discussed in the paper.
  The second  strategy for calculating the error probability is novel. It starts with the isoenergetic compression of the square well. This changes the eigenfunctions of the Hamiltonian. To keep the energy of the system unchanged, the probability weights associated with the changing eigenfunctions are dynamically modified. The reduced chamber is then probed using conventional measurement strategies and the Bayes cost calculated, which in some cases can be below the Helstrom bound. Entropy maximization associated  with the isoenergtic compression leads to an non-linearity and is the key to understand the result.




There are two main motivations for the work presented here. On the one hand it will possibly shed some light on foundational issues in quantum measurement theory, and on the other hand there are practical problems in quantum information theory, which depend on optimal state discrimination.

The structure of the paper is as follows. In section two the impact of a change of the size of the chamber is studied. In  the third section the binary choice problem between two quantum particles is tackled and the result applied to the case of a one dimensional well.
In the conclusion the result is briefly restated  and some general comments added.





 \section{Isoenergetic compression of an  one-dimensional well}
\label{sec:1b}
In this section a discussion of the isoenergetic compression of a one-dimensional well  is presented similar to an earlier paper by Bender {\it et al.}\cite{bbm2002}. The set up is simply a wave function $\chi(x)$  of a particle of mass $M$  trapped in a one-dimensional infinite square well of width $L$. The Hamiltonian is given by
\begin{eqnarray}
H= - \frac{\hbar^2}{2M}\frac{d^2}{dx^2}.\nonumber
\end{eqnarray}
At the boundary at $x=0$ and $x=L$ the wave function  vanishes.
The eigenfunctions of the Hamiltonian satisfying the boundary conditions are
\begin{eqnarray}
\bra n|x\ket= \sqrt{ \frac{2}{L}}\sin\Big(\frac{n \pi x }{L}\Big),\nonumber
\end{eqnarray}
with the respective energy eigenvalues $E_n(L)=\frac{\pi^2 \hbar^2 n^2}{2 M L^2}$.
In an earlier paper by Bender {\it et al.}\cite{bbm} infinite square wells were studied in some detail.
Any wave function satisfying the boundary conditions can be expanded into the eigenstates of the Hamiltonians of the original and  reduced chamber. The countable complex set of coefficients of the eigenstates completely describes the wave function.
The Fourier expansion before the compression is
\begin{eqnarray}
\chi(x,0)=\sum_{m=1}^{\infty} a_m \sqrt{ \frac{2}{L}}\sin\Big(\frac{m \pi x }{L}\Big) \nonumber 
\end{eqnarray}
and afterwards becomes
\begin{eqnarray}
\chi'(x,0)=\sum_{m=1}^{\infty} b_m \sqrt{ \frac{2}{L-\delta}}\sin\Big(\frac{m \pi (x ) }{L-\delta}\Big), \nonumber
\end{eqnarray}
if the size is reduced to $L-\delta$.
The final eigenstates are given by
 \begin{eqnarray}
\bra n'|x\ket=\sqrt{ \frac{2}{L-\delta}}\sin\Big(\frac{n' \pi x }{L-\delta}\Big) . \nonumber
\end{eqnarray}
For any initial vector of coefficients \begin{eqnarray}
\overrightarrow{a}= \left(
\begin{array}{l}
a_1  \\
\vdots  \\
a_n \\
\vdots
\end{array} \right)\nonumber
\end{eqnarray}
one can calculate how the isoenergetic compression transforms the coefficients into
 \begin{eqnarray}
\overrightarrow{b}= \left(
\begin{array}{l}
b_1  \\
\vdots  \\
b_n\\
\vdots
\end{array} \right).\nonumber
\end{eqnarray}
The new coefficients have to satisfy three constraints.
First the probability has to be conserved, i.e. $||\overrightarrow{b}||_2=1$.
Second the energy has to remain unchanged, since it is an isoenergetic transformation, i.e
 \begin{eqnarray}
 \sum_{n=1}^{\infty} |a_n|^2 \frac{ n^2}{ L^2}\nonumber
=\sum_{n=1}^{\infty} |b_n|^2 \frac{ n^2}{ (L-\delta)^2}.\nonumber
\end{eqnarray}
 Third the vector $\overrightarrow{b}$ has to maximize the entropy, i.e.
$ -\sum_{m=1}^{\infty} |b_m|^2 \log |b_m|^2$.
These constraints would normally not produce a unique final state,
since a complex relative phase leaves the entropy unchanged, but it narrows down the solution space.
In the end one obtains a equivalence class of optimal $\overrightarrow{b}$.

We next consider two examples to see how an isoenergetic compression effects particular states. In the first example we take a specific low energy input state such that the final vector $\overrightarrow{b}$ only has one unique non-zero component $b_1$, since the absolute phase can be neglected in quantum mechanics. This is possible through a judicious choice of of the initial probability weights and $\delta$.
This special initial state is
\begin{eqnarray}
\phi_{before}(x)&= & \frac{1}{\sqrt{2}}(|1\ket+|2\ket) \nonumber
\end{eqnarray}
and $\delta=L(1-\sqrt{2/5})$.
After the size of the well is reduced by $\delta$ the energy of the ground state increases sufficiently to produce the unique final state
\begin{eqnarray}
\phi_{after}(x)&= &|1'\ket. \nonumber
\end{eqnarray}

In the second example we choose an equal superposition of the ground state and a state of high energy $|N\ket$
\begin{eqnarray}
\psi_{before}(x)&= & \frac{1}{\sqrt{2}}(|1\ket+|N\ket). \nonumber
\end{eqnarray}
After the size of the well is again reduced by $\delta=L(1-\sqrt{2/5})$ the initial large energy of $|N\ket$ leads to a more complicated outcome with the probability spread over multiple states.
The entropy maximization is carried out in detail and a full solution is derived in section four of Bender {\it et al.}  \cite{bbm2002}.  In this paper we only provide a heuristic argument for the form of $\overrightarrow{b}$. We first consider the finite dimensional case, where  entropy is maximal, if each state has equal probability. This equal probability in the finite dimensional case can be achieved by choosing a sufficiently high initial energy state $|N\ket$. In the infinite dimensional case we know that the entropy must be higher or equal than the entropy achievable in a finite dimensional space. In addition the probability weight allowed for each component will be smaller or equal to the finite dimensional case, since an increase of the dimension can only lead to an increase in the spreading of the probability. As a result one can smear out the probability arbitrarily thinly by choosing a sufficiently large $|N\ket$ and have a probability weight for state $|1'\ket$ below any fixed number\footnote{
In addition, for finite energy initial states the probability after the compression has to asymptotically decline faster than $n^{-3}$ to avoid an infinite expectation value for the energy. As a consequence the entropy after the compression is finite as well, since $\sum_n 1/n^3 log (n)$ is finite. 
 }.
Another way of understanding the result is linked to  the non-linear nature of entropy maximization , i.e.
the compression of a superposition of two states gives a different result to the superposition of two compressed states. As an example, the initial ground state cannot be isoenergetically compressed, but in superposition with a higher energy state a compression is possible.



Before and after the contraction one can evaluate the overlap between the two test functions. This overlap changes!
The initial transition probability is $1/2$ assuming $N\geq 3$ .
During the compression the  first state simply becomes $|1'\ket$, but the second state becomes
a complicated sum of the form $\sum b_{n'} |n'\ket$, where by choosing $N$ sufficient large, the coefficient $b_1$ can be made arbitrarily small as explained above. As a result the final overlap after the compression of the wo candidate wave functions can be reduced below any chosen small number.

Before we exploit this result in the next section, let us review  three potentially controversial assumptions.
 The first assumption is to rely on an isoenergetic contraction, closely linked to an isothermal contraction, to change the size of the well.  This type of contraction was already discussed in an earlier paper by
Bender {\it et al.}  \cite{bbm2002} and seems not too unrealistic. 
The second assumption is to maximize entropy to allocate probabilities to the different eigenstates in the compression. Since a method has to be used, this seems the least arbitrary one - again see Bender {\it et al.}  \cite{bbm2002} for more details.
This leads to the third assumption of maximizing entropy in the energy basis, which assigns a preferred role to the energy basis. This is also reasonable and part of the foundation of conventional quantum statistical mechanics.
In the next section we use the preliminary results obtained so far to calculate the Bayes cost before and after the compression.

\section{Calculation of the Bayes cost before and after the compression of the chamber}

This section starts by restating the Bayesian approach to the binary decision problem as developed by Helstrom \cite{helstrom}. The result is applied to the cost calculation before and after the contraction of the square well.
The standard optimal lower bound in the binary decision problem with a $0-1$ cost, where cost $1$ is assigned to an incorrect decision and cost $0$ for a correct decision, is given by the Helstrom bound,
\begin{eqnarray}
C(\xi,|\bra\phi|\psi\ket|^2)=\frac{1}{2}-\frac{1}{2} \sqrt{1-4\xi(1-\xi)|\bra\phi|\psi\ket|^2},\nonumber
\end{eqnarray}
  for the two states $|\phi\ket$ and $|\psi\ket $, and their respective prior $\xi$ and $1-\xi$.   For a compact derivation, an example of a measurement setup for spin-1/2 particles to achieve this bound,   and  an extension  to the case of multiple copies see Brody {\it et al.} \cite{dbbm96}.

In the following paragraphs the cost is evaluated before and after  the isoenergetic compression of the chamber.
Prior to the compression the Helstrom bound for the two chosen states
\begin{eqnarray}
\phi_{before}(x)&= & \frac{1}{\sqrt{2}}(|1\ket+|2\ket), \nonumber\\
 \psi_{before}(x)&= &\frac{1}{\sqrt{2}}(|2\ket+|N\ket) \nonumber
\end{eqnarray}is
\begin{eqnarray}
\frac{1}{2}-\frac{1}{2}\sqrt{1-4\xi(1-\xi)\alpha}\nonumber
\end{eqnarray}
with the transition probability $\alpha$ equal to $1/2$.

A calculation of the cost after the compression follows next. The prior remains unchanged by the compression.
Any pair of initial states with overlap $1/2$ can be transformed prior to the isoenergetic compression into states with large energy differences without effecting the distance between the states.
 The wave function $\phi_{before}$ as a result of the compression, if $\delta$ is chosen as before, yields the new ground state.
%
It was shown in the previous section that for the second test function $\psi(x)$ the weight in $|1'\ket$ can be set equal to $\epsilon$, where $\epsilon$ is a function of $N$. Further we showed that $\epsilon$ will become smaller than any fixed number, if one chooses $N$ large enough.
The inner product of $\phi$ and $\psi$ after the compression is
\begin{eqnarray}
|\bra\phi_{after} | \psi_{after} \ket|^2=\epsilon.\nonumber
\end{eqnarray}
The new inner product between the states can therefore be reduced,
and as a result the new cost
\begin{eqnarray}
\frac{1}{2}-\frac{1}{2}\sqrt{1-4\xi(1-\xi) \epsilon}\nonumber
\end{eqnarray}
can be lower than the Helstrom bound associated with the overlap $\alpha$ before the compression and can even be arbitrarily close to $0$.

Next we calculate the difference between the original Helstrom bound and the new cost for the modified box and show  even more explicitly the violation of the Helstrom bound.
The new cost is calculated by applying the Helstrom formula to the modified state after the compression. 
The Helstrom bound can only be generally valid, if the difference between the old and new cost,
\begin{eqnarray}
C(\xi,|\bra\phi_{before}|\psi_{before}\ket|^2)-C(\xi,|\bra\phi_{after}|\psi_{after}\ket|^2)
=\frac{1}{2}-\frac{1}{2}\sqrt{1-2\xi(1-\xi)}-\Big(\frac{1}{2}-\frac{1}{2}\sqrt{1-4\xi(1-\xi) \epsilon}\Big)\nonumber
\end{eqnarray}
is always negative.
As one can readily see the difference in the cost without and with the compression is positive for  $\epsilon \leq 1/2$ for arbitrary prior $\xi$.
In other words, an increase of the energy of the second state, i.e. $N$ becomes large, leads to a reduction in the overlap after to compression and consequently to a reduction in the error probability. A simple, but suboptimal, way to find out after the compression, which of the two states has been provided, is to project the state onto the $|1'\ket$ component.
As a summary, we have provided examples where the use of an isoenergetic contraction reduces the 0-1 cost below the Helstrom bound.

\section{Conclusion}

It has been shown that the Helstrom bound in the binary quantum discrimination case can be breached.
In an intuitive way one can link the result to the non-linearity of the entropy maximization, i.e.
the compression of a superposition of two states gives a different result to the superposition of two compressed states. 

 In a companion paper to be presented at an upcoming conference potential implications for quantum algorithm will be assessed.
 What is true for states with initial overlap of $1/2$ is also true for states with arbitrary overlap.
 The important step is always to map the two candidate wave functions with arbitrary overlap into
 two states with distinct energy without initially changing the distance between the two candidate states. One will become a low energy state of the
 form
 \begin{eqnarray}
\phi(x)&= & \frac{1}{\sqrt{2}}(|1\ket+|2\ket), \nonumber
\end{eqnarray}
and the other a high energy state of the form
\begin{eqnarray}
\psi(x)&= & \alpha\frac{1}{\sqrt{2}}(|1\ket+|2\ket)+\sqrt{1-\alpha^2}|N\ket, \nonumber
\end{eqnarray}
where $\alpha$ is fixed by the overlap of the initial states. Here without influencing the core of the argument we have slightly changed the states under consideration to be able to consider any initial overlap .
The improvement in the cost beyond the Helstrom bound through the contraction of the box can be explicitly calculated as as function of the earlier introduced $\epsilon$, i.e. directly depending on $N$. As $N$ increases the ability to distinguish states with the help of an isoenergetic process will improve, i.e. $\epsilon$ is approximately  proportional to $N^{-1}$ and as a consequence inversely proportional to the square-root of the energy.


\hspace{-.38cm}The author wishes to express his gratitude to D.C. Brody for  stimulating discussion.

%

\end{document}